\documentclass[11pt]{article}
\usepackage[utf8]{inputenc}

\usepackage{bm}

\usepackage{cite}

\usepackage{nameref,hyperref}
\usepackage{eucal}
\usepackage{amsmath,amssymb}

\usepackage[right]{lineno}

\usepackage{microtype}

\usepackage{changepage}

\usepackage[aboveskip=1pt,labelfont=bf,labelsep=period,singlelinecheck=off]{caption}

\usepackage{placeins}

\usepackage{graphicx}
\usepackage{amsmath}
\usepackage[version=4]{mhchem}
\usepackage{siunitx}
\usepackage{longtable,tabularx}
\usepackage{authblk}

\usepackage{subcaption} 

\title{Smart Acoustic Lining for UHBR Technologies Engine Part 2: acoustic treatment at the intake of a scaled turbofan nacelle}

\author{E. De\;Bono\footnote{Post-doctoral researcher, LTDS \'Ecole Centrale de Lyon, emanueledeb88hotmail.it.}}

\author{M. Collet\footnote{Research Director, LTDS \'Ecole Centrale de Lyon, manuel.collet@ec-lyon.fr.}}
\author{K. Billon\footnote{Research Engineer, LTDS \'Ecole Centrale de Lyon, kevin.billon@ec-lyon.fr.}}
\affil{Univ Lyon, CNRS, \'Ecole Centrale de Lyon, LTDS, UMR5513, 69130 Ecully, France.}

\author{E. Salze\footnote{Research Engineer, LMFA \'Ecole Centrale de Lyon, edouard.salze@ec-lyon.fr.}}
\affil{Univ. Lyon, \'Ecole Centrale de Lyon, LMFA, UMR 5509, F-69134 Ecully, France.}

\author{H. Lissek\footnote{Research Director, LTS2 \'Ecole Polytechnique Fédérale de Lausanne, herve.lissek@epfl.ch.}}
\author{M. Volery\footnote{Post-doctoral researcher, LTS2 \'Ecole Polytechnique Fédérale de Lausanne, maxime.volery@epfl.ch.}}
\affil{Signal Processing Laboratory LTS2, \'Ecole Polytechnique Fédérale de Lausanne, Station 11,	CH-1015 Lausanne, Switzerland.}

\author{M. Ouisse\footnote{Professor, DMA, Université de Franche-Comté, morvan.ouisse@femto-st.fr.}}
\affil{SUPMICROTECH, Université de Franche-Comté, CNRS, institut FEMTO-ST, F-25000 Besançon, France.}

\author{J. Mardjono\footnote{Research Director, Safran Aircraft Engines, jacky.mardjono@safrangroup.com.}}
\affil{Safran Aircraft Engines, F-75015, Paris, France.}

\begin{document}

\maketitle

\begin{abstract}
The new generation of Ultra-High-By-Pass-Ratio (UHBR) turbofan engine while considerably reducing fuel consumption, threatens higher noise levels at low frequencies because of its larger diameter, lower number of blades and rotational speed. This is accompanied by a shorter nacelle, leaving less available space for acoustic treatments. In this context, a progress in the liner technology is highly demanded, prospecting alternative solutions to classic liners. The SALUTE H2020 project has taken up this challenge, proposing electro-active acoustic liners, made up of loudspeakers (actuators) and microphones (sensors). The electro-active means allow to program the surface impedance on the electroacoustic liner, but also to conceive alternative boundary laws. Test-rigs of gradually increasing complexities have allowed to raise the Technology Readiness Level (TRL) up to 3-4. In this second part, the electroacoustic liner is adapted to treat the walls at the intake of a scaled turbofan nacelle. The performance of the electroacoustic liner demonstrate its potentialities for reducing noise radiation from turbofan.
\end{abstract}

%

\section{Introduction}
The acoustic problem of interest here, is the noise transmission mitigation in an open duct, by treatment of the parietal walls with the so-called liners. Examples of industrial fields where this problem is particularly felt are the Heating and Ventilation Air-Conditioning Systems (HVAC) and the aircraft turbofan engines. The new generation of Ultra-High-By-Pass-Ratio (UHBR) turbofans, in order to comply with
the significant restrictions on fuel consumptions and pollutant emissions, present larger diameter, lower number of blades and rotational speed and a shorter nacelle. These characteristics conflict with the equally restrictive regulations on noise pollution, as the noise signature is shifted toward lower frequencies, which are much more challenging to be mitigated by parietal treatments. The acoustic liner technology applied nowadays for noise transmission attenuation at the inlet and outlet portions of turbofan engines is the so-called Single or Multi-Degree-of-Freedom liner, whose working principle relates to the quarter-wavelength resonance, and demands larger thicknesses to target lower frequencies. They are made of a closed honeycomb structure and a perforated plate which is used to provide the dissipative effect, to add mass in order to decrease the resonance frequency, and also to maintain the aerodynamic flow as smooth as possible on the internal wall of the nacelle. As the honeycomb structure is impervious, propagation is prevented transversely to the wall, therefore it can be considered as \emph{locally reacting} as long as the incident field wavelength is much larger than the size of the honeycomb cells \cite{ma2020development}.\\
A first interest for active control is the possibility to tune the resonators to different frequencies. Many adaptive Helmholtz resonator solutions have been proposed by varying either the acoustic stiffness (i.e. the cavity as in \cite{hermiller2013morphing}), or the acoustic mass (i.e. the orifice area, as in \cite{esteve2004development}), but both these techniques tended to present complex structure, excessive weight and high energy consumption \cite{ma2020development}.\\
Active Noise Cancellation (ANC) has provided alternative solutions for achieving higher attenuation levels. From the seminal idea of Olson and May \cite{olson1953electronic}, first active \emph{impedance control} strategies \cite{guicking1984active,galland2005hybrid} proposed an ``active equivalent of the quarter wavelength resonance absorber'' for normal and grazing incidence problems, respectively. The same technique was slightly modified by \cite{betgen2011new} in the attempt to reproduce the Cremer's liner optimal impedance for the first duct modes pair \cite{cremer1953theory,tester1973optimization}. As such impedance could not be achieved in a broadband sense, this approach remained limited to monotonal applications.\\
These are examples of impedance control achieved through secondary source approaches combined with passive liners, but the collocation of sensor and actuator suggested also another avenue: the modification of the actuator (loudspeaker or else) own mechano-acoustical impedance. The objective shifts from creating a ``quite zone'' at a certain location, to achieving an optimal impedance on the loudspeaker diaphragm.
This is the Electroacoustic Resonator (ER) idea, which have found various declinations, such as electrical-shunting \cite{fleming2007control}, direct-impedance control \cite{furstoss1997surface} and self-sensing \cite{leo2000self}. In order to overcome the low-flexibility drawback of electrical shunting techniques, minimize the number of sensors, meanwhile avoiding to get involved into the electrical-inductance modelling of the loudspeaker, a pressure-based current-driven architecture proved to achieve the best absorption performances in terms of both bandwidth and tunability \cite{rivet2016broadband}. It employs one or more pressure sensors (microphones) nearby the speaker, and a model-inversion digital algorithm to target the desired impedance by controlling the electrical current in the speaker coil. Compared to classical ANC strategies, the impedance control is conceived to assure the acoustical passivity of the treated boundary, and hence the stability of the control system independently of the external acoustic environment \cite{goodwin2001control}. The correlation between passivity and stability of the ER, has been analysed in \cite{de2019electroacoustic,de2022effect}. The same architecture of ER has also been exploited to conceive an alternative control algorithm capable of enforcing nonlinear acoustical responses of the ER at low excitation levels \cite{DeBono2024,MORELL2024118437,da2023experimental,morell2023control,DeBono2022spie,morellFA2023nonlinear}.
In \cite{collet2009active}, for the first time, a boundary operator involving the spatial derivative was targeted by distributed electroacoustic devices. It was the first form of the Advection Boundary Law (ABL), then implemented on ER arrays lining an acoustic waveguide in \cite{KarkarDeBono2019}, where it demonstrated non-reciprocal sound propagation. The performances of such \emph{generalized impedance} control law have been studied both in its local and nonlocal declination in \cite{de2024advection,de2023nonlocal,salze2023electro,billon2022flow,billon2021experimental,billon20222d,billon2023smart}.\\
The SALUTE projects has the objective to prompt the Technology-Readiness-Level (TRL) of a liner made up of ERs. Compact cells, each one integrating 4 microphones, a speaker and a microprocessor to execute the control algorithm, compose an array which can replace the parietal walls of a duct. In order to protect the electromechanical devices from the turbolent flow, a wiremesh, sustained by a perforated plate, is placed in front of the electroacoustic liner. Such protective liner is supposed to be quasi-transparent with respect to the acoustic field. In this second part of our contribution, we analyse the results obtained by the experimental campaign on a scaled 1:3 reproduction of turbofan engine, called PHARE-2 \cite{pereira2019new}, present in the Laboratory of Fluid Mechanics and Acoustics (LMFA) of the \'Ecole Centrale de Lyon. Two rings of 28 ER cells are employed to line a 10 cm segment at the entrance of the turbofan intake. Several engine speeds are tested, demonstrating the capability of our electro-active liner to reduce noise radiation. Before the experimental part, numerical simulations are illustrated which allowed to gain confidence about the control strategy to damp the spinning modes in a cylindrical waveguide.

\FloatBarrier
 
\subsection{Multi-modal scattering}\label{sec:multimodal scatt simulations}

\begin{figure}
	\centering
	\includegraphics[width=.8\textwidth]{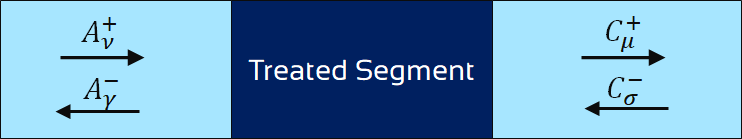}
	\caption{Illustration of the scattering problem in acoustic waveguide, with guided modes amplitudes definition on the left side, and on the right side of the acoustically treated segment.}
	\label{fig:ScatteringProblem}
\end{figure}

In this section, we aim at solving the multi-modal scattering problem in case of absence of mean-flow. It is defined as in Eq. \eqref{eq:scattering problem}, where the terms $C_\mu^+$, $A_\gamma^-$, $A_\nu^+$ and $C_\sigma^-$ are illustrated in Figure \ref{fig:ScatteringProblem}. The transmission coefficient matrix for example, $[T_{\mu,\nu}^+]$, links the amplitudes of the incident guided modes $\{ A_\nu^+ \}$ (with varying mode index ``$\nu$'') to the amplitudes of the transmitted guided modes $\{C_\mu^+\}$ (of varying index ``$\mu$'').

\begin{equation}
	\label{eq:scattering problem}
	\begin{bmatrix}
		\{C_\mu^+\}\\
		\{A_\gamma^-\}
	\end{bmatrix}
	=\begin{bmatrix}
		\bm{[T_{\mu,\nu}^+]} & \bm{[R_{\mu,\sigma}^-]} \\
		\bm{[R_{\gamma,\nu}^+]} & \bm{[T_{\gamma,\sigma}^-]}
	\end{bmatrix}
	\begin{bmatrix}
		\{A_\nu^+\}\\
		\{C_\sigma^-\}
	\end{bmatrix}
\end{equation}

\begin{figure}
	\centering
	\includegraphics[width=.6\textwidth]{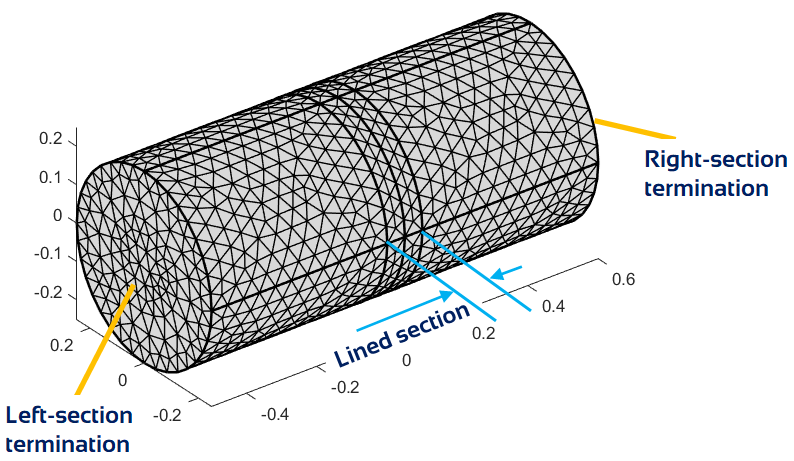}
	\caption{Mesh of the cylinder FE model, with the treated (lined) segment specified, as well as the right and left terminations.}
	\label{fig:CylinderMesh}
\end{figure}

In COMSOL Multiphysics, a Finite-Element (FE) model of a cylinder of radius $0.25$ m, with a treated segment of $0.1$ m, has been created, see Figure \ref{fig:CylinderMesh}. The segments on the left and on the right side of the treated segment has been defined sufficiently far from the respective terminations in order to avoid border effects.\\
In order to define indicators for the transmission, reflection and absorption performances of the liner, the acoustic intensity has to be defined:

\begin{equation}
	\label{eq:acoustic intensity}
	I(\omega, x) = \frac{1}{2}\int_{0}^{R} \int_{0}^{2\pi} 
	\mathrm{Re}\{ p(x,r,\theta,\omega) \; v^*(x,r,\theta,\omega) \} 
	r dr d\theta,
\end{equation}

where th superscript $^*$ indicates the complex conjugate. The acoustic intensities can be computed at the left and right terminations, by exploiting the orthogonality condition between guided modes in rigid waveguides. Supposing to excite the system by incident modes $A_\nu^+(\omega)\neq0$ and $C_\sigma^-(\omega)=0$, the pressure $p^L(\omega)$ at the left section, can be written in terms of the reflection coefficients $R_{\gamma,\nu}^+(\omega)$ as in Eq. \eqref{eq:pLeft}. The acoustic velocity $v^L(\omega)$ along the longitudinal direction, is easily retrieved in Eq. \eqref{eq:vLeft} from the Euler equation along the longitudinal direction $x$: $-\rho_0\partial_t v_x = \partial_x p$.

\begin{subequations}
	\label{eq:pLeft,vLeft}
	\begin{equation}
		\label{eq:pLeft}
		p_L(\omega) = \sum_\nu A^+_\nu(\omega)\sum_\gamma 
		J_{m(\gamma)}(k_{r_{m(\gamma) n(\gamma)}} r) e^{\mathrm{j}m(\gamma)\theta}
		\biggr( \delta_{\nu,\gamma} + R^+_{\gamma,\nu}(\omega) \biggr),
	\end{equation}
	\begin{equation}
		\label{eq:vLeft}
		v_L(\omega) = \frac{1}{\rho_0\mathrm{j}\omega}\sum_\nu (A^+_\nu(\omega))^*\sum_\gamma 
		J^*_{m(\gamma)}(k_{r_{m(\gamma) n(\gamma)}} r) e^{-\mathrm{j}m(\gamma)\theta}
		\biggr( \mathrm{j}k^*_{x_\nu}(\omega)\delta_{\nu,\gamma} - \mathrm{j}k^*_{x_\gamma}(\omega) (R^+_{\gamma,\nu}(\omega))^* \biggr),
	\end{equation}
\end{subequations}

From Eq. \eqref{eq:acoustic intensity}, and by choosing the opportune normalization of the Bessel function mode-shapes $J_{m(\gamma)}$, the acoustic intensity at the left-section termination $I_L(\omega)$ expression of Eq. \eqref{eq:I_Left} is obtained \cite{rienstra2015fundamentals}. Analogously, the acoustic intensity $I_R(\omega)$ can be evaluated at the right termination in terms of the transmission coefficient $T^+_{\mu,\nu}(\omega)$ as in Eq. \eqref{eq:I_Right}. Observe that in Eq.s \eqref{eq:acoustic intensities at terminations}, we assume to neglect the so-called \emph{tunnelling effect} \cite{rienstra2015fundamentals}, which is an acceptable assumption as long as the computation is performed sufficiently far from duct discontinuities, so that to discard contribution of evanescent modes.  

\begin{subequations}
	\label{eq:acoustic intensities at terminations}
	\begin{equation}
		\label{eq:I_Left}
		I_L(\omega) = \frac{\pi R}{\rho_0\omega}
		\sum_\nu |A^+_\nu(\omega)|^2\sum_\gamma 
		\biggr(
		\mathrm{Re}\{ k_{x_\nu}(\omega) \}\delta_{\nu,\gamma} - 
		\mathrm{Re}\{ k_{x_\gamma}(\omega) \}|R^+_{\gamma,\nu}(\omega)|^2
		\biggr)
	\end{equation}
	\begin{equation}
		\label{eq:I_Right}
		I_R(\omega) = \frac{\pi R}{\rho_0\omega}
		\sum_\nu |A^+_\nu(\omega)|^2\sum_\mu 
		\biggr( 
		\mathrm{Re}\{ k_{x_\mu}(\omega) \}|T^+_{\mu,\nu}(\omega)|^2
		\biggr)
	\end{equation}
\end{subequations}

From the expressions of $I_L(\omega)$ and $I_R(\omega)$, we can define the incident $I^+_{inc}(\omega)$, reflected $I^+_{refl}(\omega)$ and transmitted $I^+_{transm}(\omega)$ acoustic intensities, as in Eq.s \eqref{eq:incident, reflected, transmitted intensity}:

\begin{subequations}
	\label{eq:incident, reflected, transmitted intensity}
	\begin{equation}
		\label{eq:I_incident}
		I^+_{inc}(\omega) = \frac{\pi R}{\rho_0\omega}
		\sum_\nu \mathrm{Re}\{ k_{x_\nu}(\omega) \} |A^+_\nu(\omega)|^2		
	\end{equation}
	\begin{equation}
		\label{eq:I_reflected}
		I^+_{refl}(\omega) = \frac{\pi R}{\rho_0\omega}
		\sum_\nu |A^+_\nu(\omega)|^2\sum_\gamma 
		\mathrm{Re}\{ k_{x_\gamma}(\omega) \}|R^+_{\gamma,\nu}(\omega)|^2
	\end{equation}
	\begin{equation}
		\label{eq:I_transmitted}
		I^+_{transm}(\omega) = \frac{\pi R}{\rho_0\omega}
		\sum_\nu |A^+_\nu(\omega)|^2\sum_\mu 
		\mathrm{Re}\{ k_{x_\mu}(\omega) \}|T^+_{\mu,\nu}(\omega)|^2.
	\end{equation}
\end{subequations} 

As we excite the system by one single mode $\nu$ at a time, we can define the intensity reflection $r^+_{I,\nu}$, transmission $t^+_{I,\nu}$ and absorption coefficient $\alpha^+_{I,\nu}$ relative to that single mode:

\begin{subequations}
	\label{eq:single-mode reflected, transmission and absorption coeff}
	\begin{equation}
		\label{eq:rI}
		r^+_{I,\nu}(\omega) = \frac{\sum_\gamma 
			\mathrm{Re}\{ k_{x_\gamma}(\omega) \}|R^+_{\gamma,\nu}(\omega)|^2}
		{\mathrm{Re}\{ k_{x_\nu}(\omega) \}}		
	\end{equation}
	\begin{equation}
		\label{eq:tI}
		t^+_{I,\nu}(\omega) = \frac{\sum_\mu 
			\mathrm{Re}\{ k_{x_\mu}(\omega) \}|T^+_{\mu,\nu}(\omega)|^2}
		{\mathrm{Re}\{ k_{x_\nu}(\omega) \}}
	\end{equation}
	\begin{equation}
		\label{eq:alphaI}
		\alpha^+_{I,\nu}(\omega) = 1 - |r_{I,\nu}(\omega)|^2 - |t_{I,\nu}(\omega)|^2,
	\end{equation}
\end{subequations} 

Analogously, the intensity scattering coefficients $r^-_{I,\sigma}(\omega)$, $t^-_{I,\sigma}(\omega)$ and $\alpha^-_{I,\sigma}(\omega)$ for each mode $\sigma$ exciting the system from the right termination, can be retrieved after computing the scattering coefficients $R_{\mu,\sigma}^-(\omega)$ and $T_{\gamma,\sigma}^-(\omega)$.\\

After having defined the multi-modal scattering problem, we present now some interesting results concerning the attenuation of sound propagation achieved by both a purely local impedance boundary condition (BC), and the so-called Advection Boundary Law (ABL). The ABL is an extension of the local BC, where the local impedance operator defines the relationship between the boundary acceleration and a Lagrangian (not Eulerian as for purely local BCs) derivative of sound pressure, with an advection speed given in the control. This boundary law is reported below:

\begin{equation}
	\label{eq:advected B.C.}
	Z_{Loc}(\partial_t) \ast \partial_tv_n = \partial_t p + U_b\partial_{arc\theta} p \;\;\;\;\textrm{on $\partial\Omega$}.
\end{equation}

\begin{figure}
	\centering
	\includegraphics[width=.7\textwidth]{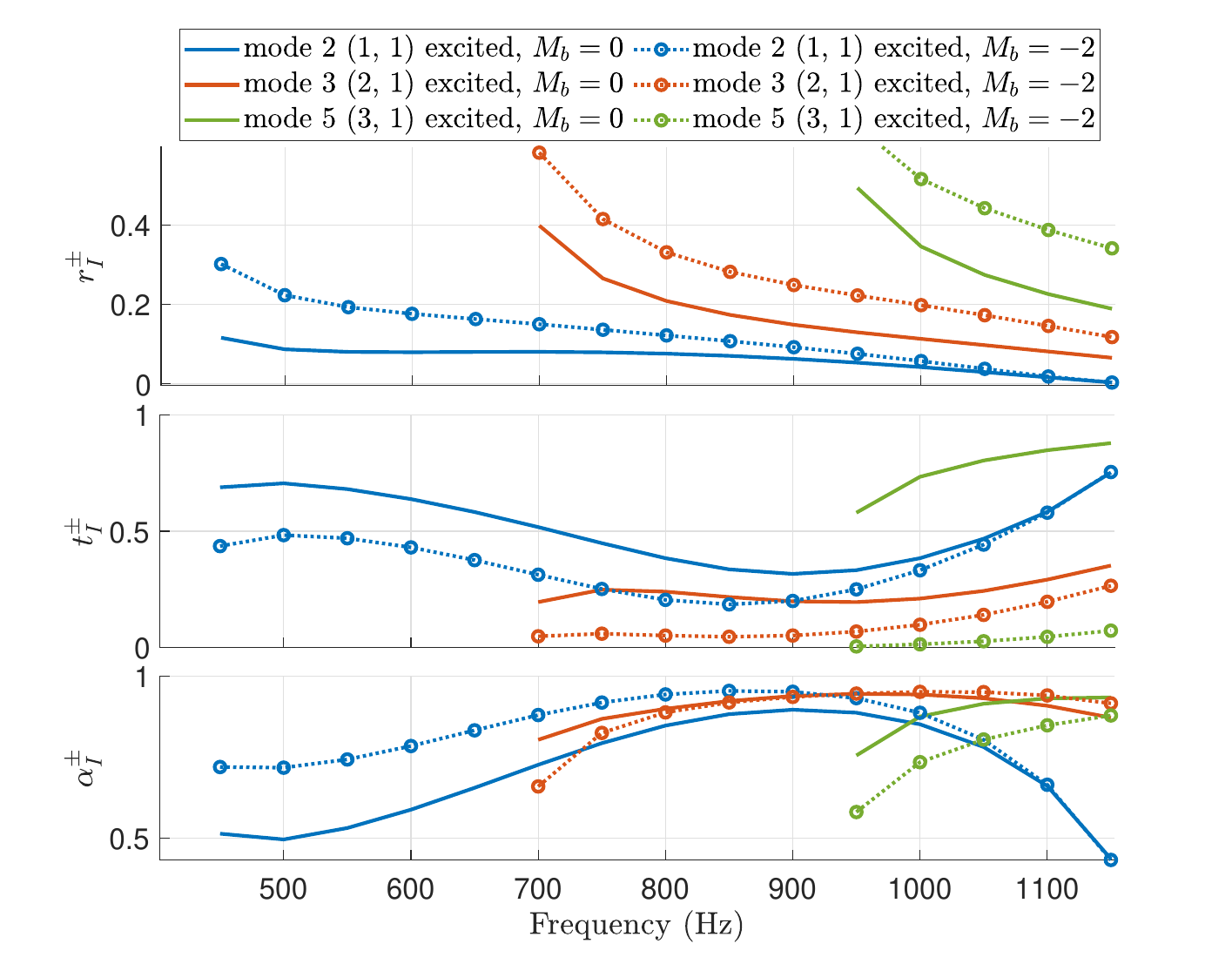}
	\caption{Intensity scattering coefficients in case of $M_b=0$ and $M_b=-2$, for the first three modes with $m\neq0$.}
	\label{fig:ScattCoeff_Mb0VSMb-2_ExcitingMode2_3_5}
\end{figure}

In Eq. \eqref{eq:advected B.C.}, $Z_{Loc}(\partial_t)$ is the differential operator in time domain corresponding to a local complex impedance, $\ast$ is the convolution operation, $v_n$ is the velocity normal to the boundary $\partial\Omega$, $p$ is the acoustic pressure, $U_b$ is the advection speed and $arc\theta=R\theta$ is the arc relative to the angle $\theta$ and radius $R$. We define $M_b=U_b/c_0$. By defining the boundary advection along the azimuthal direction, we can contrast the propagation of the spinning modes which rotate in the opposite sense with respect to $U_b$. These modes are the most interested in a turbofan nacelle.\\
We present here the simulation results obtained by such control law which has been also tested experimentally on the test-bench Phare2 \cite{salze2019new,pereira2019new}. It is hence helpful for the interpretation and appreciation of the experimental results as well.\\
In Figure \ref{fig:ScattCoeff_Mb0VSMb-2_ExcitingMode2_3_5} we compare the scattering coefficients in case of purely local target impedance (ABL with $M_b=0$), and advected law (ABL with $M_b=-2$), by exciting the first azimuthal modes with $m>0$. Observe the enhancement of the transmission attenuation in case of $M_b=-2$, corresponding to an increase in the reflection coefficient. Note also that the mode with $m=3$ cuts on around 950 Hz, and its transmission is significantly attenuated when $M_b=-2$. 


These computations have very much helped to gain confidence about the expectable results in the experimental campaign. In addition, they can provide an useful tool for \emph{optimization} of boundary control laws in multi-modal acoustic fields.

\FloatBarrier

\section{Phare test-bench}

\begin{figure}[ht!]
\centering
\includegraphics[width=0.95\textwidth]{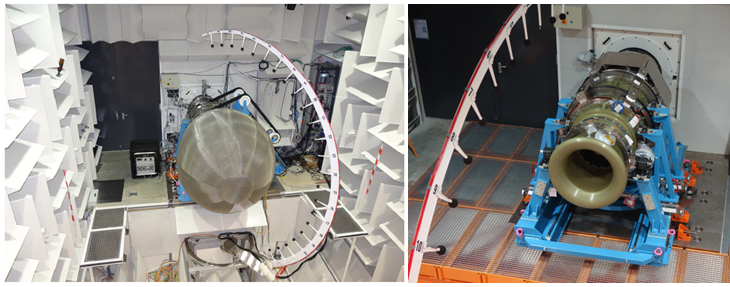}
\caption{Photos of the PHARE-2 duct stage, with (left) and without (right) turbolence screen.}
\label{fig:Phare_photo}
\end{figure}

\begin{figure}[ht!]
	\centering
	\includegraphics[width=0.7\textwidth]{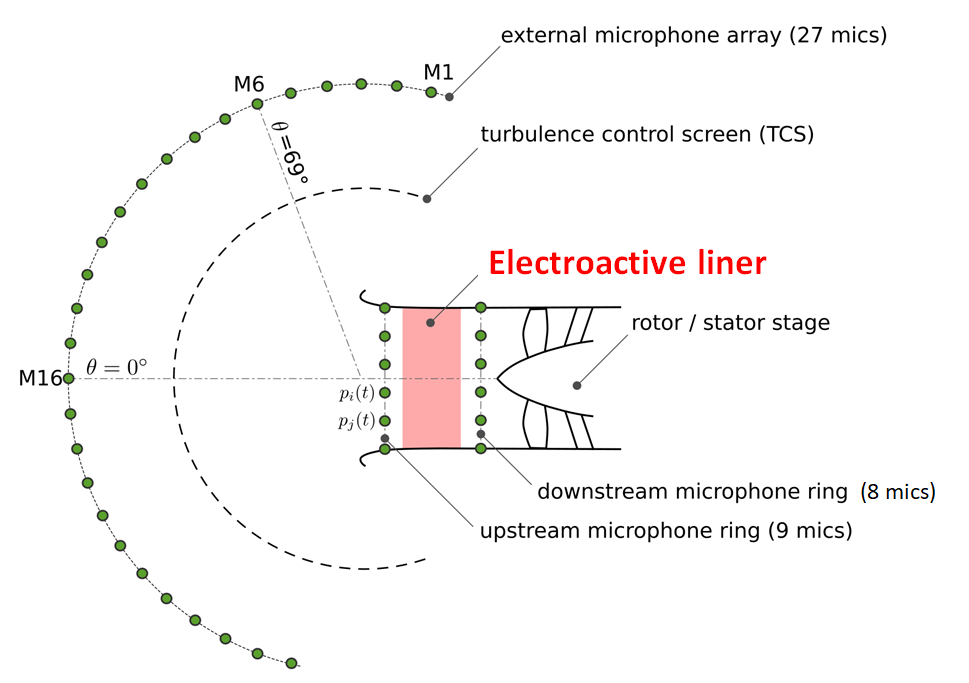}
	\caption{Sketch of the PHARE-2 duct stage.}
	\label{fig:Phare_sketch}
\end{figure}

\begin{figure}[ht!]
	\centering
	\includegraphics[width=0.95\textwidth]{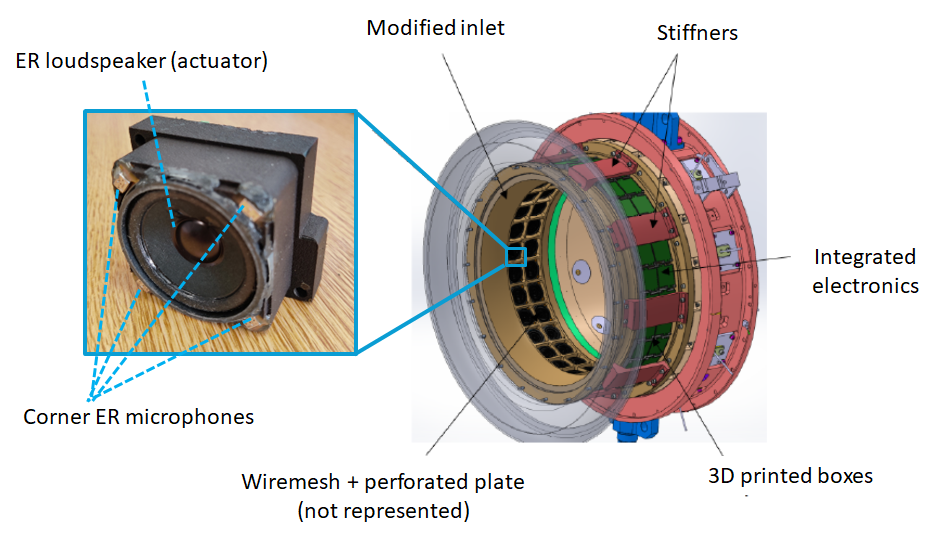}
	\caption{Photo of the single ER cell, and sketch of the modified inlet accommodating the electro-active liner.}
	\label{fig:LinedInletSketch_and_ERphoto}
\end{figure}

Here we report the results of the experimental campaign on the test-bench PHARE-2, in the Laboratory of Fluid Mechanics and Acoustics (LMFA) of the \'Ecole Centrale de Lyon, representative of a real turbofan engine
at the laboratory scale.
The turbofan engine duct is located inside an anechoic chamber. The rotational speed,
up to 16000 rotations per minute, is provided by a 3 MW
electrical engine. The air flow, up to about 40 kg/s, generated by the turbofan, passes through acoustic baffles on
the roof of the building. Using this test-rig, interest was
focused on the characterization of UHBR industrial turbofans,
regarding acoustic aspects \cite{salze2019new}, modal decomposition \cite{pereira2019new}, and aerodynamic instabilities \cite{brandstetter2019compressible}. The experimental configuration is photographed in Fig \ref{fig:Phare_photo}, and sketched in Fig. \ref{fig:Phare_sketch}. The electro-active liner is located in the inlet of the
test-rig at the location marked in red in Fig. \ref{fig:Phare_sketch}. The performances of the electro-active liner are retrieved by two microphone
rings, placed upstream and downstream the liner in the nacelle, and
by the external microphone antenna. The microphone rings in the nacelle allow to retrieve information about the azimuthal modal content downstream and upstream the liner, and the microphone antenna gives us the noise radiated upstream. The modified nacelle inlet, accommodating the ER liner is displayed in Fig. \ref{fig:LinedInletSketch_and_ERphoto} along with a photo of the single ER, made up of a loudspeaker (the actuator) and four microphones able to retrieve both the average sound pressure on the cell ($p$), and a first order finite difference approximation of the azimuthal derivative ($\partial_\{arc\theta\} p$). Those inputs are fed into a digital processor driving the electrical current in the speaker coil of each ER. A Howland current pump circuit is integrated in each cell in order to enforce the desired electrical current \cite{pease2008comprehensive}, which allows to pilot the velocity of the speaker membrane and achieve, therefore, the target (generalized) impedance operator on each ER membrane. The control algorithm is based upon the model-inversion approach, and is detailed in \cite{de2022effect,DeBono2024}.

\begin{figure}[ht!]
	\centering
	\begin{subfigure}[ht!]{0.45\textwidth}
		\centering
		\includegraphics[width=\textwidth]{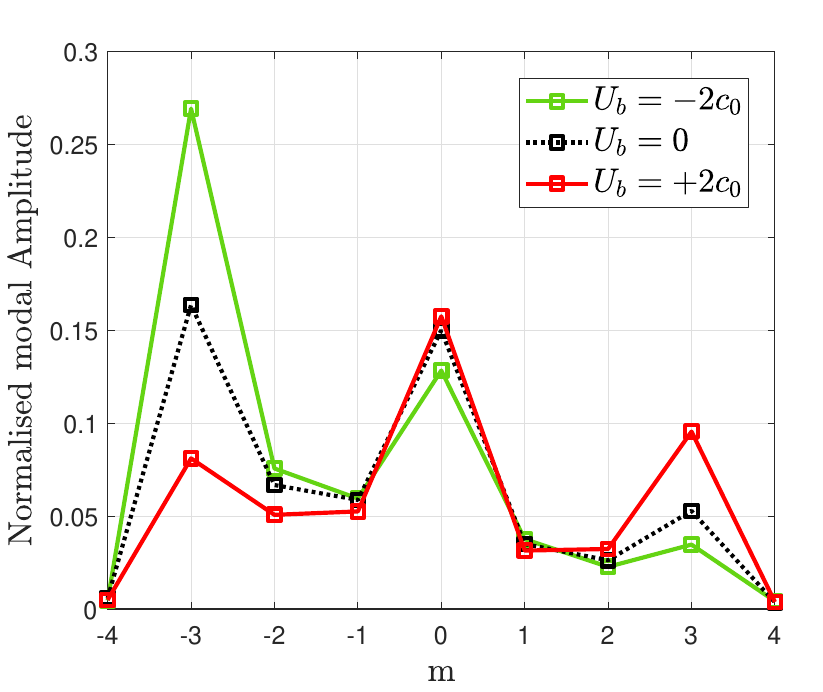}
		\centering
		\caption{}
		\label{fig:ModalContent_Point3299vs3300vs3301_EO16_upstream}
	\end{subfigure}
	\hspace{1 cm}
	\begin{subfigure}[ht!]{0.45\textwidth}
		\centering
		\includegraphics[width=\textwidth]{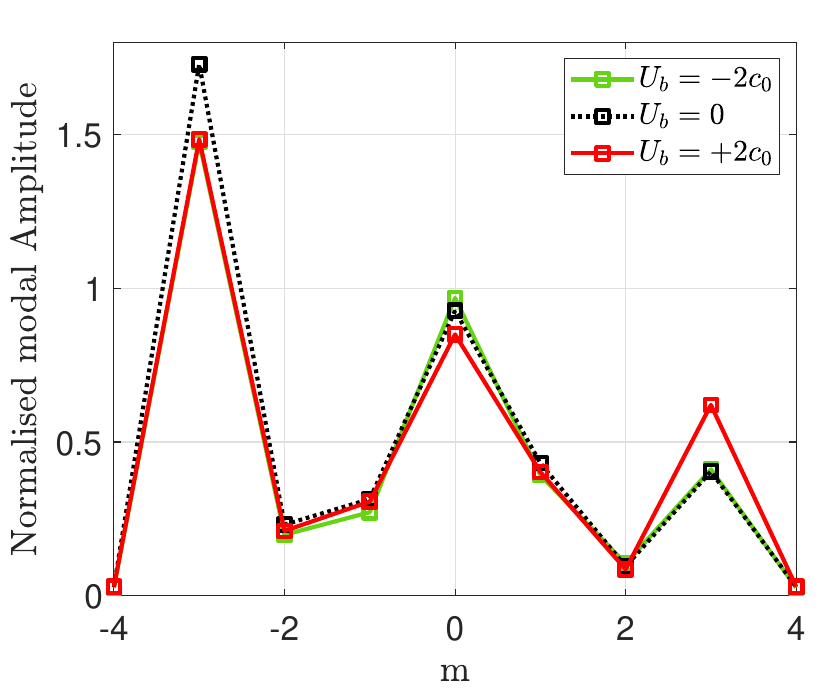}
		\caption{}
		\label{fig:ModalContent_Point3299vs3300vs3301_EO16_downstream}
	\end{subfigure}
	\caption{Normalized modal amplitude upstream (\textbf{a}) and downstream (\textbf{b}) the lined segment, around the Blade-Passing-Frequency (BPF) at $30\%$ of the nominal engine speed, with varying $U_b$.}
	\label{fig:ModalContent}
\end{figure}

\begin{figure}[ht!]
	\centering
	\includegraphics[width=0.7\textwidth]{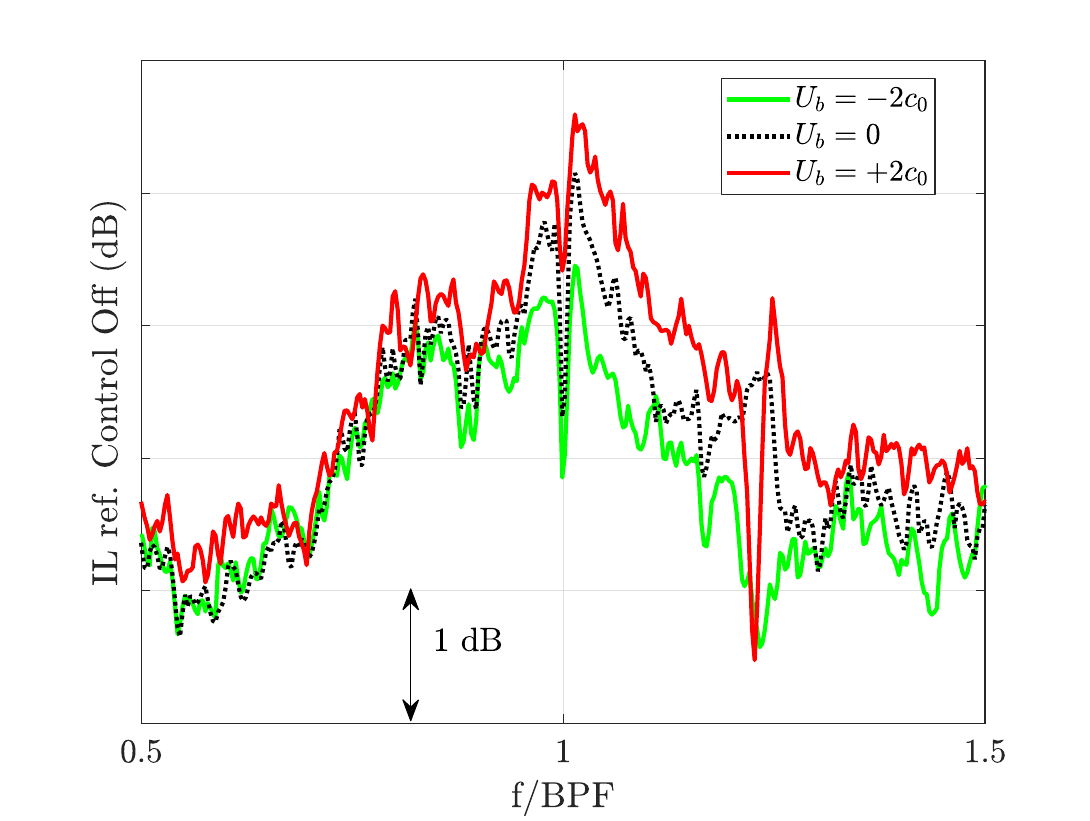}
	\caption{Insertion-Loss with respect to the Control-off configuration, at microphone 6 of the external antenna, around the BPF at $30\%$ of the nominal engine speed.}
	\label{fig:SppDelta_IL_wrtControlOff_30Nn_Interesting_28_09_Advection}
\end{figure}

The modal content upstream and downstream the lined segment can be retrieved by post-processing the measurements of the two rings of microphone as described in \cite{pereira2019new}. The normalized modal contents are reported in Fig. \ref{fig:ModalContent_Point3299vs3300vs3301_EO16_upstream} and \ref{fig:ModalContent_Point3299vs3300vs3301_EO16_downstream} 
for the upstream and downstream sections respectively, with varying $U_b$, around the Blade-Passing-Frequency (BPF) at $30\%$ of the nominal engine speed. Observe that such engine speed presents a BPF which is close to the BPF of real-size turbofans at full engine speed, therefore it is very significant to demonstrate the electro-active liner performances at this regime. Observe how for positive $U_b$, i.e. opposing the azimuthal phase speed of the acoustic modes with negative $m$, we reduce the upstream transmission of such modes, while slightly augmenting the upstream transmission of modes with positive $m$, as expected because of the non-reciprocity of our ABL \cite{KarkarDeBono2019}. Since the dominating mode $m=-3$ at the BPF is strongly attenuated in the upstream section, a reduction of noise radiation appears in Fig. \ref{fig:SppDelta_IL_wrtControlOff_30Nn_Interesting_28_09_Advection} for $U_b=-2c_0$ with respect to the purely local impedance control ($U_b=0$). In Fig. \ref{fig:SppDelta_IL_wrtControlOff_30Nn_Interesting_28_09_Advection}, indeed, we report the Insertion Loss (IL) with reference the control-off configuration, around the BPF at $30\%$ of the nominal engine speed.

\FloatBarrier

\section{Conclusions}

In this paper, we have provided the numerical simulations of an innovative nonlocal control law applied to cylindrical waveguide travelled by spinning modes. Then, we have shown some results obtained on the PHARE-2 facility, which is able to characterize the radiation from a real turbofan engine at laboratory scale (approximately 1:3 with respect to actual turbofan sizes). Two rings, each one composed by 28 Electroacoustic Resonator cells have been integrated at the inlet of the nacelle, and both Local and Advective Boundary Control laws have been tested. We have reported the performances achieved around the Blade-Passing-Frequency when the turbofan engine operates at $30\%$ of its full speed. Such configuration is significant because it corresponds to the Blade-Passing-Frequency of a real-size turbofan operating at its full speed. Our electroacoustic liner has proved robustness, stability and promising performances when faced with such a complex and harsh environment. In particular, here we have showed the enhancement of isolation (reduction of upstream transmission and radiation) achieved thanks to the implementation of an Advection Boundary Law in the azimuthal sense. The experimental performances validate the numerical predictions in terms of the potentiality of the Advection Boundary Law to attenuate spinning modes rotating in the opposite sense with respect to the artificial advection introduced on the boundary.

\section*{Acknowledgments}
The authors are particularly grateful to Benoit Paoletti,
Cedric Desbois and Antonio Pereira for their help in conducting the experiment on the PHARE-2 facility. The SALUTE project has received funding from the Clean Sky 2 Joint Undertaking under the European Union’s
Horizon 2020 research and innovation programme under grant agreement N 821093. This publication reflects
only the author’s view and the JU is not responsible for any use that may be made of the information it contains.

\FloatBarrier

\bibliography{biblio_2024}	
\bibliographystyle{unsrt}

\end{document}